# Implementation of Multiple-Step Quantized STDP Based on Novel Memristive Synapses

Yi-Fan Liu, *Student Member, IEEE*, Da-Wei Wang, *Member, IEEE*, Zhe-Kang Dong, *Senior Member, IEEE*, Hao Xie, and Wen-Sheng Zhao, *Senior Member, IEEE*

*Abstract*—Memristors have been widely studied as artificial synapses in neuromorphic circuits, due to their functional similarity with biological synapses, low operating power, and high integration density. In this work, a memristive synapse, composed of four memristors and two resistors, for SNN is designed and utilized for a neuron circuit implementing the robust spike-timing dependent plasticity learning. The synapse can be either excitatory or inhibitory by rationally arranging the resistors in the circuit. This is the first of its kind, enabling Hebbian and anti-Hebbian training without requiring additional processing of neural signals. Then, a neuron circuit is designed based on the proposed synapses. The robustness and compatibility of this neuron circuit are greatly enhanced by employing the clock-based square-wave pulsed to transmit spikes and modulate the synaptic weight. To study the performance of proposed synapses and circuit, simulations based on behavior models are carried out in the MATLAB Simulink and Simscape. Specially, a memristor model with balanced flexibility, efficiency, convergence, and emulation performance, is developed through including the nonlinear Joule effect. Using this memristor model in pattern learning, the influence of weak signal-induced weight variation on circuit performance can be rigorously assessed. This proposed circuit could give some inspiration for combining the analog memristive synapse and leaky integrate-and-fire neuron with digital control units, prompting their development as edge computing devices.

## I. INTRODUCTION

Spike-timing dependent plasticity (STDP)-based spiking neural network (SNN) is being explored to mimic the brain's capability of energy-efficient unsupervised local learning [1], [2]. The capabilities of SNNs have been extensively studied through simulations. It was found that the inherent energy efficiency and local learning feature of SNN were restricted by the "interconnect bottleneck" between separated memory and processing units [3]. A significant step in putting SNN into practice is the hardware implementation.

Commonly, synapses and neurons are core parts of the neural

This work was supported by the Natural Science Foundation of Zhejiang Province under Grants LD22F040003 and LXR22F040001, the National Natural Science Foundation of China under Grants 62101170, 62222401, and 61934006, and the Open Research Fund of State Key Laboratory of Millimeter Waves, Southeast University, under Grant K202231. (*Corresponding author: Da-Wei Wang and Wen-Sheng Zhao*).

Y. F. Liu, D. W. Wang, Z. K. Dong, and W. S. Zhao are with the with the Zhejiang Provincial Key Lab of Large-Scale Integrated Circuits Design, School of Electronics and Information, Hangzhou Dianzi University, Hangzhou 310018, China (e-mail: davidw.zoeq@gmail.com, wsh.zhao@gmail.com).

H. Xie with the School of Information and Electrical Engineering, Zhejiang University City College, Hangzhou 310015, China.

networks. Artificial synapses and neurons made up of electronic devices have been widely studied and utilized to construct the emerging SNN circuits. For artificial synapses, memristor is a promising candidate, as its functional similarity with bio-synapses, low operating power, and high integration density [4]. To date, lots of memristive synapses have been designed, including single- and multi-memristor synapses [5]-[8]. In general, the synaptic weight of the single-memristor synapse is defined by the conductance of the memristor, while that of the multi-memristor synapse is jointly determined by memristors in the circuit. Further, according to [9], synapses can be classified into three types in accordance with their polarity, namely the unidirectionally excitatory and inhibitory synapses, and the bidirectional synapse. To date, most studies were focused on excitatory and bidirectional synapses, and comparatively few research studies have been carried out on the memristive inhibitory synapse [5]-[8]. For the single-memristor synapse, to perform negative weight with a single memristor is challenging as the resistance or conductance can only be positive anyway. Although a single-memristor inhibitory synapse was realized by processing the polarity of signals sent from pre-synaptic neuron in [10], such design made all the synapses in this neuron inhibitory. Building synapses with multiple memristors is an effective means to achieve negative weight, such as the memristor bridge synapse [11] and 2M synapse [9]. However, the weights of these synapses typically range from positive to negative, corresponding to bidirectional rather than inhibitory synapses.

Apart from the polarity of artificial synapses, the signal complexity is another challenging issue for neuron circuits. Recently, various of memristive synapse-based neuromorphic circuits have been developed to implement the STDP-based SNN [12]-[19]. The defects in signal complexity, and incompatibility with conventional digital devices are restricting the practical application of memristive synapses. In practical applications, the synaptic weight is usually modulated through applying complex signals, such as biomimetic spikes [12], [15], [17], [19], to nodes of the synaptic device. The complex input signal could make the design of wave generators difficult and deteriorate neurons' robustness and compatibility with digital processing units [19].

In addition, the weight variation caused by weak signal is another factor that should be considered in circuit simulations. In most designs, similar to the read operation, the spike transmission is to read the synaptic weight using a weak signal, minimizing the influence on weight [12]-[18]. Commonly, the weak signal effect on synaptic weight is usually ignored in synapse modeling, which could critically reduce the confidence of the simulation [12], [15]-[18]. To accurately characterize the properties of artificial synapses, neurons, and SNN circuits, the weight variation caused by weak signals should be considered in modeling and simulation.

In this work, a mixed-signal SNN circuit is designed to implement the multiple-step quantized (MSQ) STDP learning [20]. To accurately characterize the nonlinearity caused by the local Joule heat and weak pulse effect in TiO$_2$ memristors, a memristor model is specially designed and utilized for general simulation. Then, a four memristors and two resistors (4M2R) synapse for SNN which is capable of either excitatory or inhibitory. Its performance is studied based on the proposed memristor model and normalized linear synaptic weight updating with soft-bound is achieved. At last, a neuron circuit requiring only square-wave pulses with a uniform amplitude is designed. This circuit can operate in a clock-synchronous manner, which improves the compatibility and expandability with digital devices.

The rest of this paper is organized as follows. In Section II, a memristor model with balanced emulation performance, flexibility, efficiency, and convergence is proposed and evaluated. Then, a 4M2R synapse for SNN is specially designed to overcome the nonlinear issue of TiO$_2$ memristor in Section III. Based on the proposed memristor model, the performance of the proposed synapse is evaluated. In Section IV, a memristor-based SNN neuron circuit is designed, and its architecture, work principle, and weak signal effect are discussed in detail. Finally, some conclusions are drawn in Section V.

## II. MEMRISTOR MODEL FOR SNN

In this section, a memristor model with balanced emulating performance, flexibility, and computational efficiency is developed to fulfill the requirements of memristive synapse simulation, ensuring both authenticity and simplicity. Until now, various of memristor models have been derived from the dopant drift model and applied to implement the performance evaluation of memristor based neuron circuits [4]. Generally, the total resistance of the memristive device is defined as

$$R_{mem} = R_{ON} \frac{w(t)}{D} + R_{OFF} \left(1 - \frac{w(t)}{D}\right) \quad (1)$$

where $D$ denotes the thickness of TiO$_2$ thin-film, $w$ is the doping depth of oxygen vacancy, $R_{ON}$ and $R_{OFF}$ are the resistances for $w(t) = D$ and $w(t) = 0$, respectively. The Ohmic characteristic of the memristive device is expressed as

$$v(t) = R_{mem} i(t). \quad (2)$$

where $v(t)$ and $i(t)$ are the biasing voltage and current, respectively. Further, the resistance switching process of the memristive device is described as

$$\frac{dw(t)}{dt} = \mu_V \frac{R_{ON}}{D} i(t) \quad (3)$$

where $dw(t)/dt$ is the switching rate and $\mu_V$ is the average dopant mobility. As shown in Fig. 1 (a), the $dw(t)/dt$ is linearly proportional to $i(t)$, that is, the model described by (3) is a linear dopant drift model.

However, this linear model is not capable of modeling artificial synapse for SNN, as it cannot characterize the nonlinear dependence of switching rate on electric current [21]. According to [27], this switching nonlinearity is mainly caused by the local Joule heating effect. The device self-heating can result in temperature increase which accelerates the thermally activated drift of oxygen vacancy and leads to rise in the switching rate. To characterize this phenomenon, the dependence of $dw(t)/dt$ on $i(t)$ or $v(t)$ can be described by sinh [21], [28], exponential [12], or power [18], [29] function. Considering the nonlinear dependence on $i(t)$, (3) can be modified as

$$\frac{dw(t)}{dt} = \mu_V \frac{R_{ON}}{D} g(i) \quad (4)$$

where $g(i)$ is the Joule function.

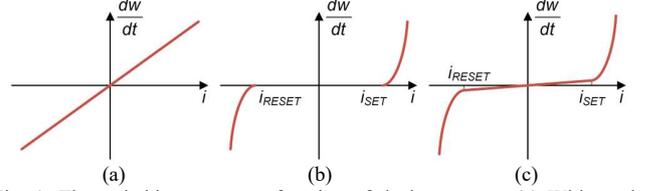

Fig. 1. The switching rate as a function of device current. (a) Without the nonlinearity and weak signal effect, (b) with the nonlinearity but without weak signal effect, and (c) with both the nonlinearity and weak signal effects.

Further, as shown in Fig. 1(b), the switching rate is usually assumed to be zero when the amplitude of the input current or voltage was below the SET/RESET thresholds in prior works [12], [15]-[18], [28], [29]. Actually, the resistance variation is not zero when the input is below the thresholds, as shown in Fig. 1(c), and this phenomenon should be considered in simulations to ensure greater realism and prevent potential design issues. Consequently, similar to the TEAM model, a continuous Joule function is proposed to include the power relationship [28].

$$g(i) = a_0 \left(\frac{i(t)}{i_0}\right)^{2q-1} \quad (5)$$

where $q$ is an integer positive exponent, $i_0$ and $a_0$ are fitting parameters. In addition, the boundary effect is also included by multiplying a window function $f(w)$ [4], (4) is rewritten as

$$\frac{dw(t)}{dt} = \mu_V \frac{R_{ON}}{D} g(i) f(w) \quad (6)$$

where $f(w)$ can be any off-the-rack window function, such as Strukov [4], Joglekar [22], Prodromakis [23], Biolek [24], and Zha [25], which characterize the dependence of $dw(t)/dt$ on $w$. In this way, a framework for a memristor model that meet the requirements of artificial synapse simulation for SNN is obtained.

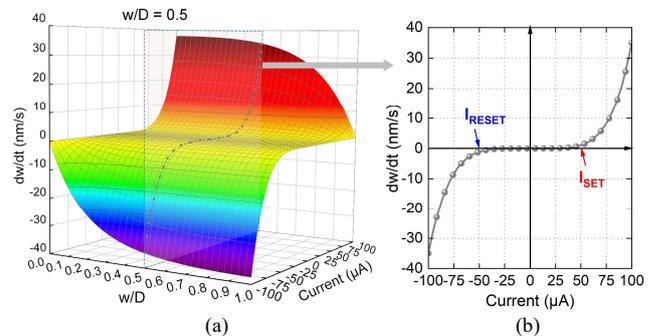

Fig. 2. (a) Switching rate $dw(t)/dt$ as functions of w/D and device current. (b) Switching rate as a function of device current at w/D = 0.5.

The performance of the proposed memristor model is verified through a typical simulation analysis. In the following simulations, parameter values listed in TABLE I are utilized, and the Zha window function with parameters $p = 10$, $j = 1$, and $D = 10$ nm, is chosen as it aggregates the virtues of several prior window functions and is computationally efficient. The switching rate described by equations (5) - (8) is illustrated in Fig. 2(a). For the ON switching process, corresponding to the positive device current, the absolute value of switching rate decreases with $w/D$, while the opposite trend is observed for the OFF-switching process. This is mainly contributed by the boundary effect. Fig. 2(b) shows the power relationship between the switching rate and device current, with the initial state of $w_0 = 5$ nm. Further, the weak signal effect is also considered in this model, namely the switching rate is non-zero for device current lower than the threshold.

TABLE I
THE VALUE OF PARAMETERS UTILIZED IN THE SIMULATIONS

| Param. | $R_{ON}$ ($\Omega$) | $R_{ON}$ ($k\Omega$) | $\mu_v$ (m²s⁻¹V⁻¹) | $a_0$ (A) | $i_0$ (mA) | $q$ |
|---|---|---|---|---|---|---|
| Value | 100 | 16.0 | $10^{-14}$ | 40 | 1.0 | 3.0 |

The pinched hysteresis loop of the memristor model, driven by a sinusoidal input with a frequency of 10 Hz and an amplitude of 1 V, is illustrated in Fig. 3(a). Obvious nonlinearity is observed when the boundary between the doped and undoped layers approaches ends of the device. The hard switch of memristor can be triggered with a signal having sufficient high amplitude and low frequency. Fig. 3(b) shows the continuous hard switch excited by a 2V, 1 Hz input signal. The above simulations are carried out using the Simulink environment, and the memristor model is implemented in Simscape language.

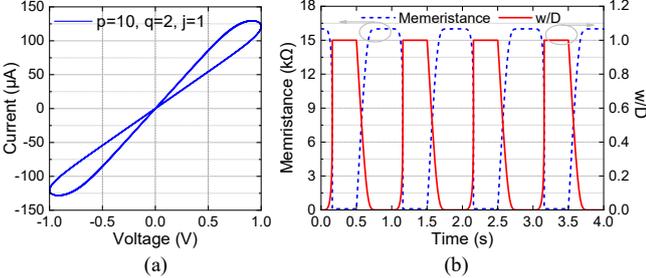

Fig. 3. (a) The pinched hysteresis loop of the in-house memristor model, driven by the sinusoidal voltage signal $v(t) = \sin(20\pi t)$. (b) Continuous hard switch of the memristance and normalized dopant width $w/D$, driven by a sinusoidal voltage signal $v(t) = 2\sin(2\pi t)$.

Compared with the extensively known TEAM [28] and VTEAM [29] models, the proposed model is free from nested sinusoidal or exponential functions, which makes it more computationally efficient and convergent. In comparison with the simplified memristor models, the proposed model characterizes more factors that could influence the SNN circuit functionality and performance, making the simulation more informative. The aim of this model is to drive more efficient and effective SNN circuit simulation and design.

## III. 4M2R SYNAPSE FOR SNN

In this section, an 4M2R artificial synapse, composed of four memristors, two resistors and a differential amplifier, is designed for SNN circuits. As shown in Fig. 4, the synapse can be either excitatory or inhibitory, depending on the positions of the resistors in the circuit. According to [33], the excitatory synapse and inhibitory synapse correspond to synapses with only positive and negative weights, respectively. The proposed synapse is derived from the memristor bridge synapse [11], which has been utilized in analog artificial neural networks (ANNs) [30], [31]. The Its capability of presenting bi-polar weight and linear weight modulation are also desirable features for SNN synapses. However, some adjustments have to be made before it can be applied to SNN circuits, as the working principles of SNN and ANN are different in some ways.

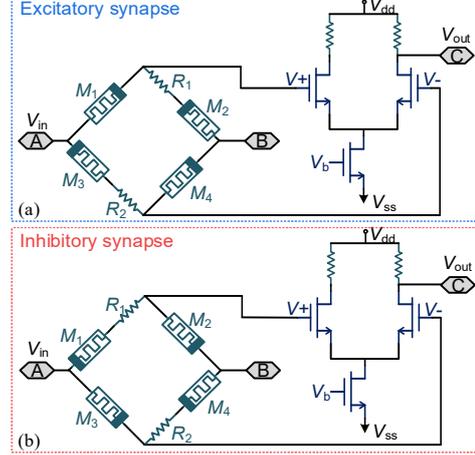

Fig. 4. Schematic diagram of the proposed 4M2R synapse for SNN. (a) excitatory synapse and (b) inhibitory synapse. The synaptic weight is presented by 4 memristors and 2 resistors on the left. The differential amplifier working at its linear region, acts as a subtractor.

The proposed synapse possesses several outstanding merits in comparison with the conventional memristor bridge synapse. First of all, the proposed 4M2R synapse can be either excitatory or inhibitory. Different from the synapses used in ANN circuit which are usually ambipolar, the synapses utilized in SNN circuits are required to be excitatory or inhibitory, to emulate the biological synapses that release different neurotransmitters. As shown in Fig. 4(a), the excitatory artificial synapse can be designed by connecting a resistor in series to memristors $M_2$ and $M_3$, respectively. The resistance of resistors should be identical to the maximum resistance of the memristor. According to the circuit, the weighted output voltage $V_{out}$ is defined to be the difference between node voltages $V_+$ and $V_-$, and it is obtained by a differential amplifier with magnification $A$. When a positive spike $V_{in}$ is applied at Terminal A, with Terminal B grounded, $V_{out}$ can be calculated as

$$V_{out} = A(V_+ - V_-)$$
$$= A\left(\frac{M_2 + R_1}{M_1 + M_2 + R_1} - \frac{M_4}{M_3 + R_2 + M_4}\right)V_{in} \quad (7)$$

For simplification, (10) can be rewritten as

$$V_{out} = \psi \times V_{in} \quad (8)$$

and $\psi$ is the synaptic weight and it is expressed as

$$\psi = A\left(\frac{M_2 + R_1}{M_1 + M_2 + R_1} - \frac{M_4}{M_3 + R_2 + M_4}\right)$$

where $R_1 = R_2 = R_{OFF} = 16$ $k\Omega$ and $A = 1.1$. Accordingly, the variation range of $\psi$ is calculated to be [0.0312, 0.9938], positive only. The resistance of resistors is chosen according to the maximum resistance of memristor, as shown in Fig. 5(a). The value of $A$ is defined to normalize the variation range of the synaptic weight to be [0, 1], as shown in Fig. 5(b). Following a similar design method, the inhibitory synapse can be obtained. The difference is that the polarity of four memristors is inverted and the resistors are connected in series to memristors $M_1$ and $M_4$, respectively, as shown in Fig. 4(b). This circuit can ensure that the weight variation range is negative only. That is, with the same programming signal in Fig. 4(c), the direction of weight modulation is opposite to that of the excitatory synapse. Using this inhibitory synapse, the anti-Hebbian learning rule [34] can be implemented without manipulating the neuron signals.

Secondly, another merit is that the weight of the 4M2R synapse can be calculated individually by the differential amplifier operating in its linear region. This is mainly due to that different from the ANN, the traditional activation functions are not required in the SNN [30], [31]. Further, the voltage output signal of the designed synapse makes all signals transmitted between neurons are voltage signals, which could greatly facilitate the design of auxiliary facilities, such as analog multiplexers (MUXs) and allow for a more flexible connection between the neurons.

Generally, artificial synapses involve two operations, the weight modulation and spike transmission. In this work, strong signals with an amplitude of 4 V are utilized to modulate the synaptic weight while weak signals with 2 V amplitude are utilized to transmit spikes. The signal amplitudes are mainly determined by the SET/RESET thresholds of the memristor and can be adjusted accordingly. The memristance tunning of $M_1$, $M_2$, $M_3$, and $M_4$ and the corresponding weight modulation resulted from applying the potentiation and depression programming signal in Fig. 5 (c), are shown in Fig. 5(a) and Fig. 5(b), respectively. It is observed that both the change rates of memristance and synaptic weight gradually decrease to zero as they approach their value range boundary. According to [11], this is mainly caused by the boundary effect of memristors. Although, the nonlinearity in synaptic weight modulation can seriously influence the performance of ANN, the artificial neuromorphic network is not sensitive to it. This is mainly due to that such nonlinearity is also ubiquitous in the biological synapses which is known as the soft-bound [32].

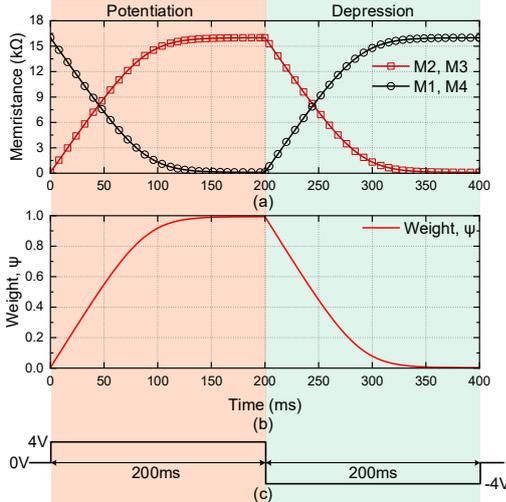

Fig. 5. Potentiation and depression characteristic of the excitatory 4M2R synapse. The initial memristance are $M_1(0) = M_4(0) = 16 \; k\Omega$ and $M_2(0) = M_3(0) = 100 \; \Omega$. (a) $M_1(t)$, $M_2(t)$, $M_3(t)$ and $M_4(t)$. (b) Synaptic weight $\psi(t)$. (c) Programming voltage pulse across Terminals A and B.

Similarly, the transmitted weak signals can also potentiate or depress the weight to some extent. For instance, with the initial weight $\psi_0 = 0.5$, the weight increments by weak and strong signals with a duration of 10 ms are $\Delta\psi_{weak} = 0.0024$ and $\Delta\psi_{strong} = 0.07441$, respectively. The ratio $\Delta\psi_{weak}/\Delta\psi_{strong}$ is about $3.225\%$, which indicates that there is significant differentiation between the effects of these two types of signals. Although weak signals can also potentiate or depress the weight slightly, the effect is small enough to avoid causing functional issues. The influence of weak signals on circuit performance is discussed in the following section.

## IV. MIXED-SIGNAL ARTIFICIAL NEURON

Commonly, a biological neuron is composed of several core components, they are the soma, dendrite, synapse, axon hillock, and axon [33], as shown in Fig. 6(a). The dendrites work as the receivers for the neuron, and they receive the signals from other neurons through the connected synapses. Then, the received signals are transmitted into the soma, where these signals are processed and integrated through updating membrane potential. The membrane potential increases in case of a potentiation input signal received and decreases in case of the depression input signal. Normally, when the inputs are weak or less frequent, a sub-threshold membrane boost will leak out gradually, whereas when the inputs are strong or frequent enough, the membrane potential can be boosted to exceed the firing threshold. Then, the neuron fires and generates an action potential (or a spike) in the axon hillock, which carries massages and is transmitted to other neurons through the axon. The mixed-signal artificial neuron proposed in this work follows a similar rule to biological synapses.

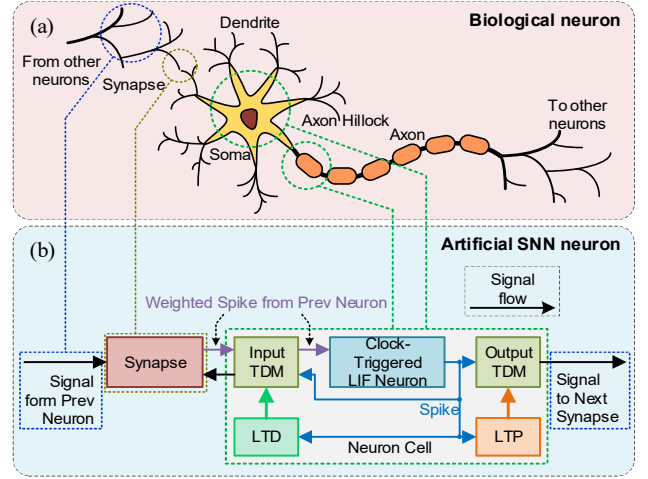

Fig. 6. Analogy between a biological neuron and an artificial SNN neuron. (a) Schematic diagram of a biological neuron. (b) Block diagram of our proposed artificial SNN neuron.

### A. Working Principle and Read-Write Dilemma

Fig. 6(b) shows the block diagram of the proposed artificial SNN neuron. The circuit modules in this diagram are defined based on their functions corresponding to the biological neuron. The synaptic weight is stored in the artificial synapse and the spikes received from other neurons are weighted and then output through the Terminal C of synapse, as shown in Fig. 4. The analog membrane potential integration, leakage, as well as synchronous digital spike emission are realized by a clock-triggered leaky integrate-and-fire (LIF) neuron module. The long-term potentiation (LTP) and long-term depression (LTD) modules are used to store the timing information of the previous spikes of the neuron, known as the trace, which is the reference to synaptic weight plasticity [36]. Based on this artificial neuron circuit, the symmetric interpretation scheme of nearest-neighbor pair-based STDP is implemented [36], [37].

The read-write dilemma is a typical issue in the memristive neuromorphic circuit design [15]. Biologically, the inference and learning are executed through coordinated biochemical processes, and they can function simultaneously without interfering with each other. However, in the memristive artificial synapse, the spike transmission and weight modulation share the same input port, which leads to mutual interference. To address this issue, a time-division multiplexing (TDM) scheme is implemented, eliminating the interference by executing spike transmission and weight modulation at different time slots. To implement the TDM scheme, the time is discretized into isochronous steps which is called frame as shown in Fig. 7(a). Each frame is composed of three timeslots, they are timeslot 0 for spike transmission, timeslot 1 for potentiating programming signal, and timeslot 2 for depressing programming signals. The signals in Timeslots 0 and 1 are received from the pre-synaptic neuron (Pre), while the signal in Timeslot 2 is sent by the Post-synaptic neuron (Post). The clock tree of the neuron circuit is shown in Fig. 7(b), and a clock frequency of 100 Hz is set in this work. Practically, the clock signal can be generated by an external source and synchronized with the global clock when the neuromorphic circuit is applied to edge computing. A fractional-3 frequency is utilized for the clock-triggered LIF neuron. In the simulation model, ideal models for ancillary devices are utilized, including operational amplifiers (OPAs), analog multiplexers (MUXs), differential amplifiers, NMOS, comparators, and digital devices.

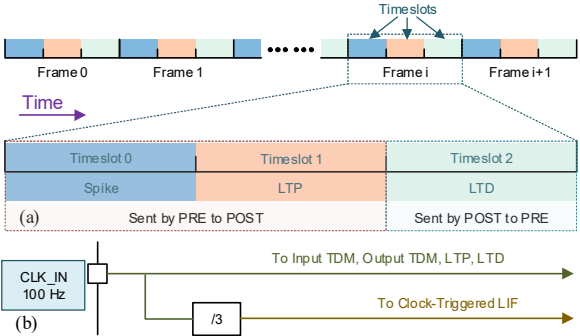

Fig. 7. (a) TDM scheme of signal through synapses. (b) Clock tree of the artificial neuron circuit.

### B. Clock-Triggered LIF Neuron

The schematic of the proposed clock-triggered LIF neuron circuit is shown in Fig. 8. This neuron consists of an inverting integrator with multiple inputs to integrate the received spikes, a comparator to present the firing threshold, and a trigger-controlled voltage source to generate output spikes. Following the working principle will be described in conjunction with the exemplary sequence shown in Fig. 9. The spikes with 2 V amplitude are received and weighted by the synapse first and then the weighted signal shown in Fig. 9(a) is transmitted into Post. Subsequently, the received pulses are converted into currents through the input resistors $R_{in,j}$ (j =1, 2, …, n, n is the total number of pulses). After that, the currents are collected through the virtual ground at the negative input node of the inverting integrator, resulting in the increment of the capacitor voltage $V_{MP}$ as shown in Fig. 9(b). The $V_{MP}$ represents the membrane potential of the artificial neuron. The firing threshold is set through applying a bias voltage with an amplitude of $V_{th}$ to the positive node of the comparator. The $V_{th}$ is set to be -0.45 V in this experiment as the gain of inverting integrator is negative. The values of the other lump components are $R_{in,j} = 100\ k\Omega$, $R_{ref} = 900\ k\Omega$, and $C = 1\ \mu F$.

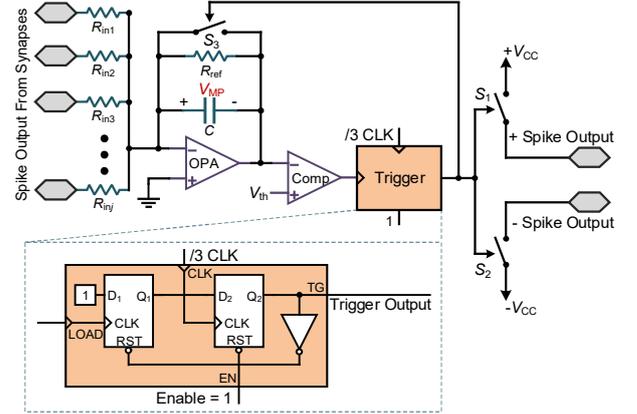

Fig. 8. Schematic of the Clock-triggered LIF module.

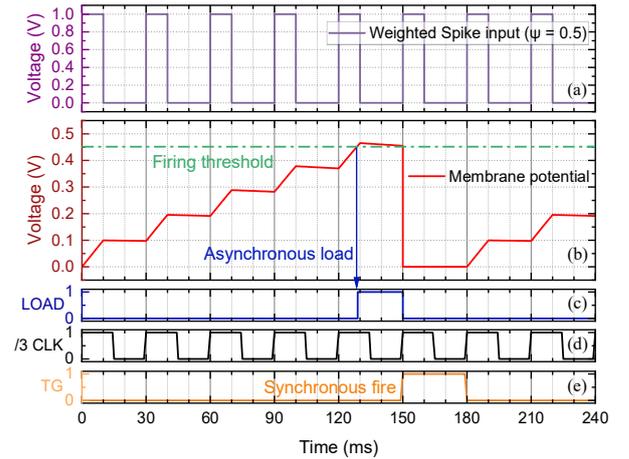

Fig. 9. Time sequence of the clock-triggered LIF neuron. (a) Weighted spikes with $\psi = 0.5$, (b) membrane potential $V_{MP}$ integration and leakage, (c) The output of comparator, (d) clock, and (e) synchronous fire.

The trigger in the LIF module is composed of two D flip-flops and a NOT gate, as shown in the inset of Fig. 8. As $V_{MP}$ accumulating above the $V_{th}$, $Q_1$ is set to high-level at the rising edge of the comparator output pulse asynchronously, presenting the "loaded" state of the trigger, as shown in Fig. 9(c). Upon the rising edge of the fractional-3 clock (Fig. 9(d)), $Q_2$ is set to high-level to closed switches $S_1$, $S_2$, and $S_3$, by which the synchronous fire is touched off (Fig. 9(e)). As a result, a positive spike and a negative spike are emitted through $S_1$ and $S_2$, respectively. The charge on the capacitor $C$ is released through $S_3$, namely, $V_{MP}$ is reset to zero. Meanwhile, the trigger is reset by the inverter and $Q_1$ is set to low-level. Upon the next rising edge of the fractional-3 clock, $Q_2$ is set to low-level and the output of the trigger is over. In this way, a pair of bi-polar spikes with an amplitude $V_{CC} = 2$ V is generated, last for a whole frame. Each neuron can be enabled or disabled individually through the enable (EN) port of the trigger and by setting the EN port to low-level, the corresponding neuron is unable to fire.

## C. Implementation of STDP

STDP is a timing-dependent specialization of the Hebbian learning rule, which is derived from biological neuron behavior [38]. Fig. 10 shows the schematic diagram of a typical SNN fragment in which the Pre is connected to the Post through a synapse. According to the working principle of pair based STDP, the synaptic efficiency should be strengthened if the Pre and Post spikes are causally related, otherwise, the synaptic efficiency should be weakened. The amount of the weight variation is proportional to the value of the traces left by the neuron spikes [36].

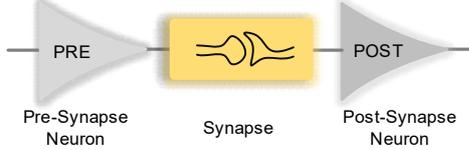

Fig. 10. The schematic diagram of a typical SNN fragment.

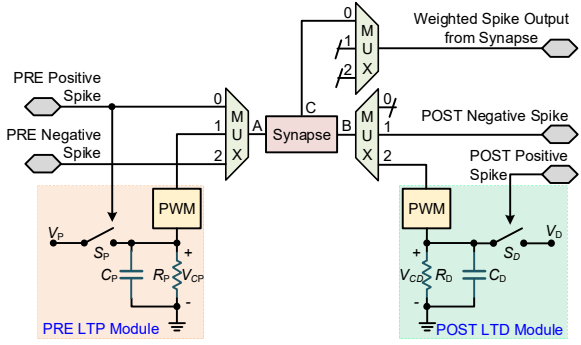

Fig. 11. The schematic circuit diagram for a SNN fragment, implemented based on the proposed synapse and neurons. Pre and Post traces are realized by the RC circuits in Pre LTP and Post LTD modules, respectively. Signals transmitted through busses are selected by timeslot using the analog multiplexers with M = 3.

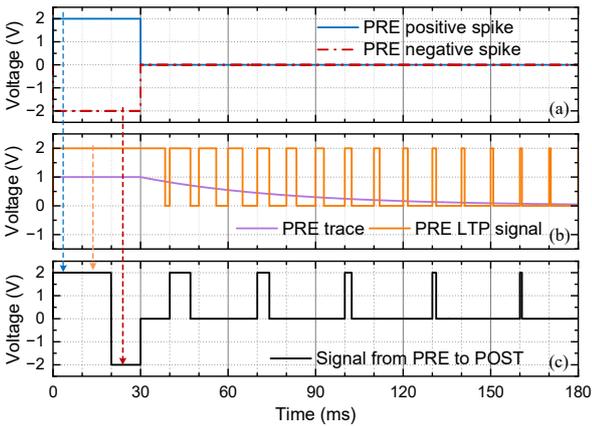

Fig. 12. Constitution of signals sent from Pre to Post.

The schematic circuit diagram for a SNN fragment implemented with the proposed artificial synapse and neuron is illustrated in Fig. 11. In this design, MUXs are employed to select the signals to be emitted or received in turn by timeslot, controlled by a shared $M = 3$ up counter. In the Pre synapse neuron, port 0 and 2 of the MUX are connected to input signals and port 1 is connected to the output of a pulse width modulation (PWM). When a positive-negative spike pair is generated by the LIF module in one frame, as shown in Fig. 12(a), the $S_P$ is switched on by the positive spike and the $C_P$ in the Pre LTP module is charged to an initial electric potential $V_P$, namely, $V_{CP} = V_P$. The output of the PWM stay high during the first frame due to the switch is always on, as shown in Fig. 12(b). At the same time, as shown in Fig. 12(c), the MUX drives the positive spike, output of PWM, and negative spike to the synapse during timeslot 0, 1, and 2, sequentially. By the end of the first frame, the Pre spike is over and the $S_P$ is switched off. Then, the $C_P$ starts discharging through the $R_P$, resulting in an exponential drop in the capacitor voltage $V_{CP}$ as shown in Fig. 12(b). Here, the $V_{CP}$ is utilized to characterizes the trace of Pre spike due to their similar updating and decreasing dynamics [36]. The amplitude of the trace, namely $V_{CP}$, is sampled at the beginning of timeslot 1 of each frame and encoded into a pulse with the width proportional to sampled trace amplitude by the PWM. In this way, the signal sent from Pre to Post is obtained, as shown in Fig. 12(c).

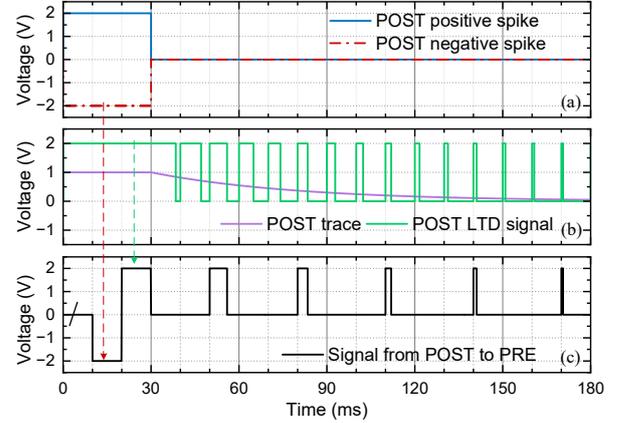

Fig. 13. Constitution of signals sent from Post to Pre.

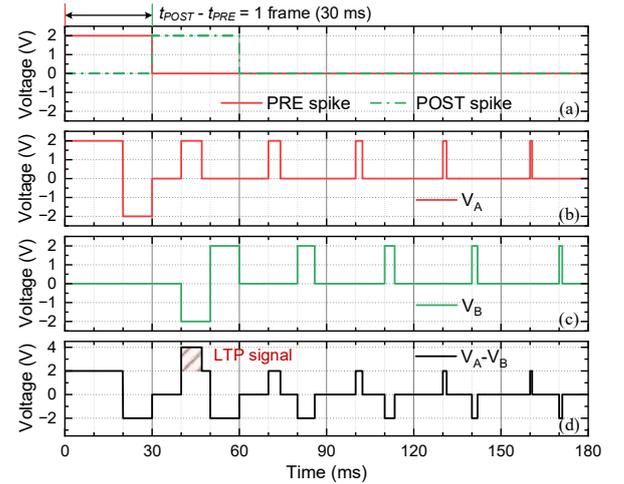

Fig. 14. The generation process of LTP signal by Pre-Post spike pair. The increment of the synaptic weight depends on $\Delta t = t_{Post} - t_{Pre}$, namely, width of the LTP signal.

For the signal backpropagated from Post to Pre, there are some differences in the working principle. Like the Pre synapse neuron, a positive-negative spike pair is also generated by the LIF module in the Post one, as shown in Fig. 13(a). However, the positive spike generated by the Post travels forward to the Posterior neuron only, thus no signal is backpropagated during the timeslot 0. Consequently, port 0 of the MUX in the Post synapse neuron is grounded as shown in Fig. 11. During the timeslot 1, the generated negative spike is driven to port B. The output of Post LTD module is connected to port 2 of the MUX and driven to port B during timeslot 2. Similarly, the positive spike is utilized to control the switch $S_D$ and the trace of Post spike is characterized, sampled, and encoded in the same way, as shown in Fig. 13(b). In this way, the composite signal sent from Post to Pre is obtained and applied to Terminal B of the synapse, as shown in Fig. 13(c).

Fig. 14 demonstrates the generation process of LTP signal by the Pre-Post spike pair. As shown in Fig. 14(a), a Pre spike is generated in the first frame while a Post spike is generated in the second frame. In timeslot 1 of the first frame, the Pre LTP signal is applied to port A of the synapse while the port B is virtually grounded. That is a weak signal is applied to the synapse and its weight could be rarely affected. In the second frame, the Post spikes and port B of synapse are connected to negative voltage with -2 V amplitude, namely a LTP signal with 4 V amplitude is applied to the synapse in timeslot 1, as shown in Fig. 14(d). According to section III, the synaptic weight could be adjusted by this LTP signal.

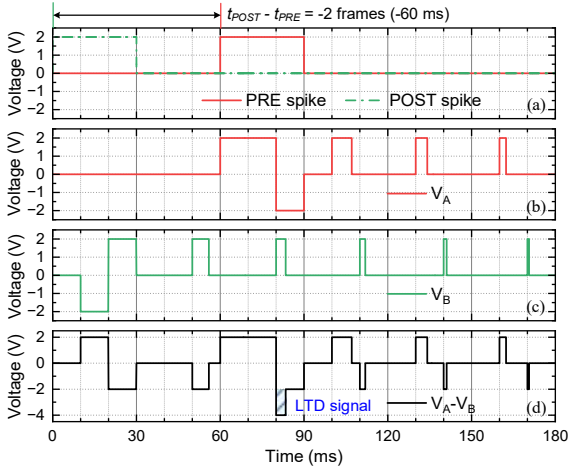

Fig. 15. The generation process of LTD signal by Post-Pre spike pair. The increment of the synaptic weight depends on $\Delta t = t_{\text{pre}} - t_{\text{post}}$, namely, the width of LTP signal.

Fig. 15 demonstrates the generation process of long-term depression (LTD) signal by the Post-Pre spike pair. As shown in Fig. Fig. 15(a), the Post spikes in the first frame and the Pre spikes in the third frame. In Fig. 15(b) and (c), the Pre LTP and Post LTD signals generated by the Pre and Post are illustrated, respectively. It can be observed that, during the first two frames, the Post LTD signal is applied to the port B of synapse while the port A is virtually grounded. That is, only weak signals are applied to the synapse which can rarely influence the synaptic weight. In the third frame, the Pre spikes and port B of synapse are connected to voltage with -2 V, namely an LTD signal with -4 V amplitude is applied to the synapse in timeslot 1, as shown in Fig. 19(d). Similarly, the synaptic weight could be adjusted by this LTD signal.

Overall, the adjustment range of synaptic weight is determined by the time interval $|\Delta t|$ between the Pre and Post spikes. For instance, $|\Delta t|$ in Fig. 15 is 2 frames while that in Fig.

14 is 1 frame, and thus, the width of the LTD signal is observed to be much narrower than that of the LTP signal. In general, the LTP or LTD signal width decreases exponentially in accordance with $|\Delta t|$. As stated in Section III, in the linear weight updating region, $|\Delta \psi|$ is almost proportional to the duration of the programming signal. As a result, $|\Delta \psi|$ decreases exponentially with respect to $|\Delta t|$, which enables the realization of an MSQ Hebbian STDP window, as depicted in Fig. 16(a). Moreover, when Pre and Post spike simultaneously, the LTP and LTD programming signals appear in the same frame with the same width, which would offset each other, causing the weight variation $|\Delta \psi| = 0$. For the inhibitory synapse, the weight modulation follows the anti-Hebbian learning rule, as depicted in Fig. 16(b).

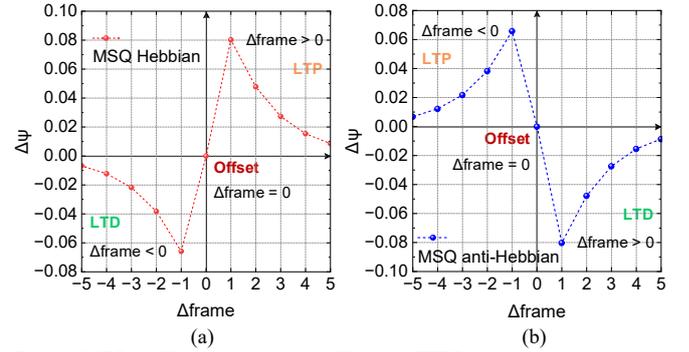

Fig. 16. MSQ (a) Hebbian and (b) anti-Hebbian STDP window achieved by the proposed synapse and weight modulation scheme. The initial weight is $\psi_0 = 0.5$, and the $\Delta \psi$ is induced by one pair of spikes at each point.

Following, the practical applicability of the proposed design scheme is evaluated through using the state-of-art memristor model. An exemplary study is carried out by utilizing the VTEAM memristor to construct the 4M2R synapse [39], [40]. Accordingly, the resistances of resistors and the magnification of digital amplifier are adjusted to $R_1 = R_2 = R_{\text{OFF}} = 8 \ k\Omega$ and $A = 1.7$. The global clock frequency of the neuron circuits is set as 1 kHz (3 ms for 1 frame) to match the time scale of the STDP window experimentally measured on biological synapses [41], and the LTP and LTD modules are adjusted to fit the magnitude and dropping speed of the target STDP window. The simulation results show excellent fitting with the experimentally measured STDP learning window, as shown in Fig. 17(a). Additionally, the corresponding anti-Hebbian learning window is also illustrated in Fig. 17(b).

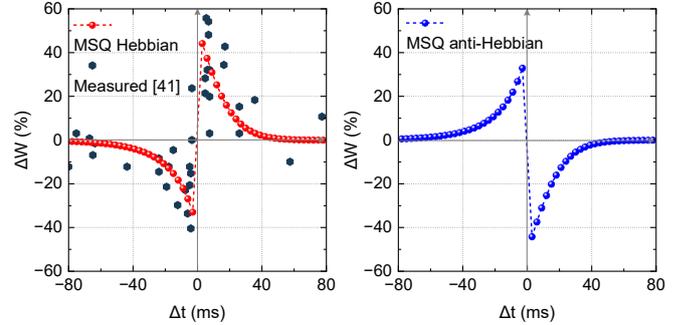

Fig. 17. Implementation desired STDP window with VTEAM memristor. (a) Comparison of the simulated and measured STDP Hebbian learning window and (b) and the corresponding anti-Hebbian learning window.

Different from achieving through reversing the polarity of the neuronal output signal in previous work [10], the anti-

Hebbian STDP is achieved directly by the four memristor with reversed direction in the inhibitory synapse in this work. In turn, it is possible to stimulate and modulate the excitatory or inhibitory synapse using the unified form of signals, which would greatly improve the design flexibility of the neural network circuits. Combining specially designed neuron topologies, this proposed design scheme is promising in applications such as field programmable SNN devices.

### D. Pattern Learning by MSQ STDP

To estimate performance of the proposed MSQ STDP window in weight modulation, a network with 2 layers is designed and simulated. The first layer of the $3 \times 3$ Pre array serves as the input neurons and they are numbered according to the inset in Fig. 18. The second layer is composed of Posts, and they correspond one-to-one with the Pres in the first layer. A phrase of the Pre spike sequence is illustrated in Fig. 18, where the Pres with No. 1, 3,4 ,6 and 8 spike simultaneously in the first frame to mark a "V" pattern. These simultaneous spikes tend to induce a Post spike in the following frame in case of the $V_{MP}$ exceeding $V_{th}$. Then the Pres with No. 2, 5, 7, and 9 spike subsequently in the 3rd, 4th, 5th, and 6th frames to imitate the input noise. The repetition frequency of spike group (one epoch) shown in Fig. 18 is 3.33Hz, namely its time interval between the first Pre spikes of adjacent groups, is $1/\rho = 300m$.

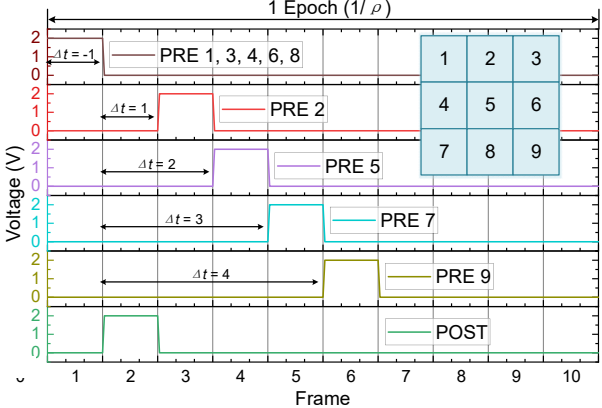

Fig. 18. An epoch of the Pres and Post spikes. The repetition frequency $\rho = 3.33$ Hz, namely each epoch lasts for $1/\rho = 300$ ms (or 10 frames).

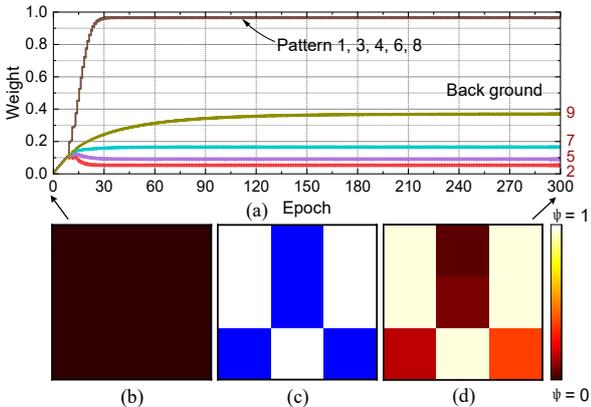

Fig. 19. Pattern learning by initializing the synaptic weights to be 0.

The evolution of synaptic weights for the first training is illustrated in Fig. 19(a). Before training, all the synaptic weights are initialized to be 0 through applying a negative programming signal with a duration of 1 s, as shown in Fig. 19(b). Then the Pre spikes described in Fig. 18, corresponding to the pattern depicted in Fig. 19(c), are applied periodically. It is observed that no Post spike is emitted in the first 10 epochs, which is mainly due to that the synaptic weights are low and $V_{th}$ is not reached. Ideally, the long-term synaptic plasticity should be unachievable in the absence of Post spikes. However, the synaptic weights are observed growing gradually in the first 10 epochs. This is mainly caused by the weak signal effect, including the transmitted spike signal in timeslot 0 and the Pre LTP signal in timeslot 1. The Post begins to emit spikes in the second frame of each epoch just after the $V_{th}$ is reached, then LTP and LTD occur in synapses with pattern input and noise input, respectively. The strength of LTD is inversely proportional to $\Delta t$ between noise Pre spike and Post spike. As shown in Fig. 19(a), the weights of the 7th and 9th synapses grow along with time, and this is mainly because that the weak signal effect is stronger than LTD. After about 300 epochs, the weight potentiation and depression reach an equivalent state. The final synaptic weights distribution is presented in Fig. 19(d), where the weights of synapses with pattern input are potentiated to high levels and the synaptic weights correlated to noise input converge to a lower level.

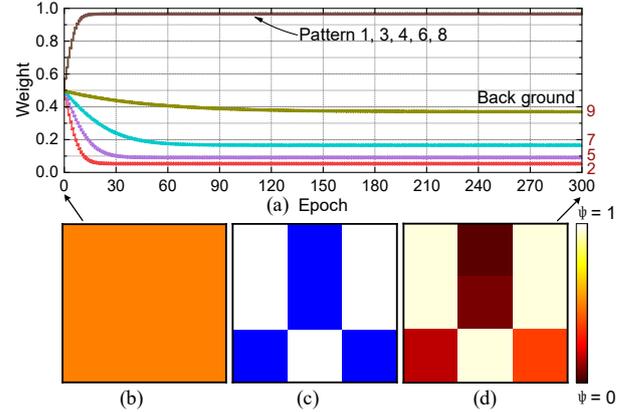

Fig. 20. Pattern learning by initializing the synaptic weights to the middle of the weight's dynamic range.

In the second round of training, all the synaptic weights are initialized to the midpoint of the weight range, as shown in Fig. 20(b). According to [19], this initialization approach could facilitate the training process and thanks to the linearity of the synaptic weight, the initialization can be easily implemented for the proposed artificial synapse. Again, the Pre spikes shown in Fig. 18 are utilized for training. Comparing the synaptic weight curves in Fig. 19(a) and Fig. 20(a), it is observed that the convergence rate is significantly improved by optimizing the initial state. Fig. 20(d) shows the synaptic weight distributions after 300 epochs which achieves stability and is identical to that shown in Fig. 19(d). Ideally, only the programming signal can modulate the synaptic weights, therefore, the synapses with pattern input signals should be modulated to the highest weight level while those with noise input signals are the lowest. However, the results show that the weights of synapses with noise inputs converge to certain levels during the train tests. This is caused by the combined effects of weak signal effect and LTD signal, and the maximum adjustments are determined by the Post-Pre spike pair intervals $|\Delta t|$.

## V. Conclusions

In this work, a 4M2R memristive synapse which was capable of being either excitatory or inhibitory, was designed and utilized for neuron circuit to implement the robust STDP learning. The performance of proposed synapse was studied through simulations, and the results indicated that linear synaptic weight modulation could be achieved through properly designing the synapse circuit, even only the simple memristive devices with nonlinear conductance tuning were utilized. Then, a mixed-signal artificial neuron was designed based on the proposed synapse, in which both spike transmission and weight modulation were realized using clock synchronous square-wave pluses with uniform amplitude. The robustness, feasibility and compatibility with conventional digital devices are enhanced noticeably. After that, the MSQ STDP was realized by regulating the duration of the programming signal, which was encoded through implementing the PWM. Compared with previous work, the proposed scheme can achieve anti-Hebbian STDP without manipulating neural signals, resulting in better flexibility in neuronal synaptic assembly.

To demonstrate the performance of the proposed SNN circuit, a $3\times3$ pattern learning was carried out and the results indicated that the MSQ STDP was successfully implemented and the weak signal effect on synaptic weight was included through using the improved memristor model. The nonideality in the training process and final weight distribution reveals that the weight variation caused by weak signal can compromise the circuit's performance. This phenomenon can be considered in future designs to achieve greater realism and preciseness, while avoiding potential design issues. We hope our model, circuit and discovery can give an inspiration for the development of memristor-based SNN circuits as edge computing devices.